\documentstyle[nato,amssymb]{crckapb}

\begin{opening}
\title{Hybrid Inflation and Supergravity}
\author{C. Panagiotakopoulos}
\institute{Physics Division, School of Technology, Aristotle University of\\
Thessaloniki, 54006 Thessaloniki, Greece}
\end{opening}
\runningtitle{Hybrid Inflation and Supergravity}

\begin{document}

\begin{abstract}
Hybrid inflation is a natural scenario in the absence of supersymmetry.
In the context of supergravity, however, it has to face the naturalness
problems of the initial conditions and of the adequate suppression of the
inflaton mass. Both can be successfully addressed in a class of
models involving K\"ahler potentials associated with products of
$SU(1,1)/U(1)$ K\"ahler manifolds and ``decoupled'' fields
acquiring large vacuum expectation values through D-terms. 
\end{abstract}

Inflation offers an elegant solution to many cosmological problems
\cite{linde90}. However, ``natural'' realizations of the inflationary scenario
are hard to find. ``New'' and ``chaotic'' inflation \cite{linde90} invoke a very
weakly coupled scalar field, the inflaton, in order to reproduce the observed
temperature fluctuations $\frac{\Delta T}{T}$ \cite{cobe}
in the cosmic background radiation (CBR). To overcome this
naturalness problem Linde proposed the ``hybrid'' inflationary scenario
\cite{hyb,lyth} involving a coupled system of (two) scalar fields which
manages to produce the temperature fluctuations in the CBR with natural values
of the coupling constants. This is achieved by exploiting the smallness
in Planck scale units ($M_{P}/\sqrt{8\pi }\simeq 2.435515\times
10^{18}\>GeV=1$ which are adopted throughout our disscusion)  
of the false vacuum energy density associated with the phase transition
leading to the spontaneous breaking of a symmetry in the post-Planck era.
In the case that the broken symmetry is a gauge symmetry one of
the (two) scalar fields involved is not a gauge singlet.

Linde's potential is given by 
\begin{equation}
V(\varphi ,\sigma )=(-\mu ^{2}+\frac{1}{4}\lambda \varphi ^{2})^{2}+\frac{1}{
4}\lambda _{1}\varphi ^{2}\sigma ^{2}+\frac{1}{2}\beta \mu ^{4}\sigma ^{2},
\end{equation}
where $\varphi ,\sigma $ are real scalar fields, $\mu $ is a mass parameter
related to the symmetry breaking scale and $\lambda ,\lambda _{1},\beta $
are real positive constants. Notice that at $\sigma ^{2}=\sigma _{c}^{2}=2
\frac{\lambda }{\lambda _{1}}\mu ^{2}$ the $\sigma $-dependent mass-squared
of $\varphi ,$ $m_{\varphi }^{2}(\sigma )=-\lambda\mu ^{2} +\frac{1}{2}
\lambda _{1}\sigma ^{2},$ vanishes. Then, for $\sigma ^{2}>\sigma _{c}^{2}$,
$m_{\varphi }^{2}(\sigma )>0$ and the potential at fixed $\sigma $ as a
function of $\varphi $, namely $V_{\sigma }(\varphi )$, has a minimum at $
\varphi =0$ with $V_{\sigma }(0)=\mu ^{4}(1+\frac{1}{2}\beta \sigma ^{2}).$
For $\sigma ^{2}<\sigma _{c}^{2}$ instead, $m_{\varphi }^{2}(\sigma )<0$ and 
$V_{\sigma }(\varphi )$ has a minimum at $\left| \frac{\varphi }{2}\right|
=\left( -\frac{m_{\varphi }^{2}(\sigma )}{\lambda ^{2}}\right) ^{\frac{1}{2}
}.$ Moreover, $\left| \frac{\varphi }{2}\right| =\frac{\mu }{\sqrt{\lambda }}
,\sigma =0$ minimizes $V(\varphi ,\sigma ).$

Let us assume that $\frac{2}{\beta }$ $\gg \sigma ^{2}>$ $\sigma _{c}^{2},$
$\varphi =0$ and $\beta \ll 1$.
Then, the potential is dominated by the almost constant
false vacuum energy density, i.e.
$V(0,\sigma )=\mu ^{4}(1+\frac{1}{2}\beta \sigma ^{2}) \simeq \mu ^{4},$
the ``slow-roll'' parameters $\epsilon ,\left|\eta\right| \ll 1,$
since $\epsilon \equiv \frac{1}{2}\left(\frac{V^{^{\prime }}}{V}\right) ^{2}
=\frac{1}{2}\beta ^{2}\sigma ^{2}\ll
\beta ,$ $\eta \equiv \frac{V^{^{\prime \prime }}}{V} =\beta ,$
and the universe experiences an inflationary stage with Hubble parameter
$H\simeq \frac{\mu ^{2}}{\sqrt{3}}$. During inflation the motion of the inflaton
field $\sigma $ is governed, in the ``slow-roll'' approximation, by the
equation
\begin{equation}
\frac{d\sigma }{dt}\simeq-\frac{1}{\sqrt{3}}\beta \mu ^{2}\sigma .
\end{equation}
Inflation ends at $\sigma \simeq \sigma_{c}$ with a rapid phase transition
towards the true minimum 
$\left| \frac{\varphi }{2}\right| =\frac{\mu }{\sqrt{\lambda }}, \sigma =0$.   
The number of e-foldings for the cosmic time interval $(t_{in},t_{f})$,
corresponding to a variation of $\sigma $ between the
values $\sigma _{in}$ and $\sigma _{f}$ (with $\sigma _{in}^{2}>\sigma
_{f}^{2}$), is 
\begin{equation}
\int_{t_{in}}^{t_{f}}Hdt \simeq \beta ^{-1}\ln \frac{\sigma _{in}}{\sigma _{f}}
=N(\sigma _{in})-N(\sigma _{f})
\end{equation}
with $N(\sigma )\equiv \beta ^{-1}\ln \frac{\sigma }{\sigma _{c}}.$  Also the
spectral index of density fluctuations $n\simeq1+2\beta $
is almost scale invariant and slightly larger than $1$ for $\beta \ll 1 .$ 
Assuming, as it turns out to be the case for $\beta \ll 1$, that the measured
(quadrupole) anisotropy $\frac{\Delta T}{T}\simeq 6.6\times 10^{-6}$
is dominated by its scalar component
$\left( \frac{\Delta T}{T}\right) _{S}\simeq \left( 12\pi \sqrt{5}
V^{^{\prime }}\right) ^{-1}V^{\frac{3}{2}}$ (evaluated at $\sigma =\sigma
_{H}=\sigma _{c}e^{\beta N_{H}}$, where $N_{H}\equiv N(\sigma _{H})\simeq
50-60$ is the number of e-foldings of ``observable''
inflation) and choosing $\lambda _{1}=\lambda ^{2}$ we have 
\begin{equation}
\frac{\mu }{\sqrt{\lambda }}=12\pi \sqrt{10}\left( \frac{\Delta T}{T}\right)
_{S}\frac{\beta }{\lambda }e^{\beta N_{H}}\simeq 0.79\times 10^{-3}\frac{
\beta }{\lambda }e^{\beta N_{H}}.
\end{equation}
Taking $\beta \sim 10^{-4},\ \lambda \sim 1$ we obtain an intermediate scale
of symmetry breaking $\frac{\mu }{\sqrt{\lambda }}\sim 10^{-7}$ and an
electroweak-scale inflaton mass $\sqrt{\beta }\mu ^{2}\sim 10^{-16}$.
Taking, instead, larger values of $\beta $ we obtain larger scales. For
example, $\beta \simeq 1/35,\ \lambda \simeq 10^{-2}, N_{H}\simeq 55$ 
gives $\frac{\mu }{%
\sqrt{\lambda }}\simeq 1.1\times 10^{-2}$ and $\mu \simeq 1.1\times 10^{-3}.$

At this point we should remark that $\mu $ cannot be arbitrarily large since
there is an upper bound on the energy density scale $V_{infl}^{\frac{1}{4}
}\simeq V_{\sigma _{H}}^{\frac{1}{4}}\simeq \mu $ where the ``observable''
inflation begins. By exploiting the fact that the tensor component $
\left( \frac{\Delta T}{T}\right) _{T}$ of $\frac{\Delta T}{T}$ satisfies $
\left( \frac{\Delta T}{T}\right) _{T}^{2}\simeq \left( 720\pi ^{2}\right)
^{-1}6.9V_{\sigma _{H}}<\left( \frac{\Delta T}{T}\right) ^{2}$ we
immediately derive the bound 
\begin{equation}
V_{infl}^{\frac{1}{4}}\simeq V_{\sigma _{H}}^{\frac{1}{4}}\simeq \mu
\lesssim 1.46\times 10^{-2}.
\end{equation}

How natural are the initial conditions that lead to the hybrid inflationary
scenario \cite{in}? We assume that the energy density 
$\rho$ of the universe is dominated by $V(\varphi ,\sigma )$.
Let us start away from the inflationary trajectory and choose the energy
density $\rho _{0}$ to satisfy the relation $\mu ^{4}\ll \rho _{0}\lesssim 1.$
Moreover, we assume that $\varphi ^{2}$ starts somewhat below $\sigma ^{2}.$
Then, the relevant term in $V$ for our discussion is the term
$\frac{1}{4}\lambda ^{2}\varphi ^{2}\sigma^{2}.$
We would like $\varphi $ to oscillate from the beginning as a
massive field due to its coupling to $\sigma $ and quickly become very close
to zero. In contrast $\sigma ^{2}$ should stay considerably larger than
$\sigma _{c}^{2}$. Thus, for $\mu ^{4}\ll \rho \leq \rho _{0}\lesssim 1$ it is
required that $\frac{4}{9}\frac{m_{\varphi }^{2}}{H^{2}}\gg 1\gg \frac{4}{9}
\frac{m_{\sigma }^{2}}{H^{2}}$ or $\varphi ^{2}\ll \frac{8}{3}\ll \sigma
^{2}.$ When $\rho \sim \mu ^{4}$, instead, $\left|\sigma\right|$ remains larger
than $\left|\sigma_{c}\right|$ provided 
$\frac{4}{3}\frac{m_{\sigma }^{2}}{\mu 4}\lesssim 1$
or $\varphi ^{2}\lesssim \frac{3}{2}\frac{\mu ^{4}}{\lambda ^{2}}$.
If we allow $\left| \sigma _{0}\right| \gg 1$, $\left|
\varphi _{0}\right| $ does not have to be very small. For example, with the
choice $\beta \simeq 1/35,\ \lambda \simeq 10^{-2}$, $\mu \simeq 1.1\times
10^{-3}$ we could have $\left| \sigma _{0}\right| \simeq 4.5$, $\left|
\varphi _{0}\right| \simeq 1$. If, instead, we insist that $\left| \sigma
_{0}\right| <1$ we are forced to start very close to the inflationary
trajectory $\rho_{0} \simeq \mu^{4} \ll 1$ and severely fine tune the starting
field configuration ($\left| \sigma_{c}\right| \ll \left| \sigma_{0}\right|<1, $
$\left|\varphi _{0}\right| \lesssim \frac{\mu ^{2}}{\lambda }\ll 1$).

This severe fine tuning becomes more disturbing since the
field configuration at the assumed onset of inflation,
where $H=H_{infl}$, should be homogeneous over dinstances
$\sim H_{infl}^{-1}$. Notice that $H_{infl}^{-1}$ is larger than the Hubble
distance at the end of the Planck era ($\rho =$ $\rho _{in}\simeq 1$) as
expanded (according to the expansion law $R\sim \rho ^{-\frac{1}{3\gamma }},$
where $R$ is the scale factor of the universe) till the assumed onset of
inflation (at $\rho =\rho _{_{infl}}$) by a factor $\frac{H_{infl}^{-1}}{
H_{in}^{-1}}\left( \frac{\rho _{infl}}{\rho _{in}}\right) ^{\frac{1}{3\gamma }
}=\left( \frac{\rho _{infl}}{\rho _{in}}\right) ^{-\frac{3\gamma -2}{6\gamma }
}\gg 1,\ $if$\ \ \gamma \gtrsim 1.$ Therefore, in order for any
inflation to start at an energy density scale $\rho _{infl}^{\frac{1
}{4}}\simeq $ $V_{infl}^{\frac{1}{4}}\ll 1$, the initial field configuration
at $\rho =\rho _{in}\simeq 1$ (where initial conditions should be set)
must be very homogeneous over distances $\sim \left( V_{infl}^{-\frac{1}{4}
}\right) ^{2\frac{3\gamma -2}{3\gamma }}\gg 1$. Such a homogeneity is hard to
understand unless a short period of inflation took place at
$\rho \sim 1$ \cite{double} with a number of e-foldings
$\gtrsim 2\frac{3\gamma -2}{3\gamma }\ln \left(V_{infl}^{-\frac{1}{4}}\right)$.
An early inflationary stage might also eliminate the requirement of severe fine
tuning of the field configuration at $\rho=\rho_{in}$ since,
in addition to the homogenization of space, it could alter the dynamics
during the early stages of the evolution of the universe.

An inflation taking place at an energy density $\rho_{1}\gg\rho_{infl}$, however,
although eliminates existing inhomogeneities it generates new ones due to
quantum fluctuations. These fluctuations are $\sim \frac{H_{1}}{2\pi }$ for
massless fields and generate inhomogeneities over distances $\sim H_{1}^{-1}$
resulting in a gradient energy density $\sim \frac{H_{1}^{4}}{4\pi ^{2}}=\frac{
\rho_{1} ^{2}}{36\pi ^{2}}$ which falls with the expansion only like
$R^{-2}\sim \rho ^{\frac{2}{3\gamma }}.$
The size of this gradient energy density when $\rho $ falls to $\rho
_{infl}\simeq V_{infl}$ should be smaller than $V_{infl}.$
This gives an upper bound on the energy density $\rho_{1}$
(towards the end) of the first stage of inflation
\begin{equation}
\rho _{1}\lesssim \left( 6\pi \right) ^{\frac{3\gamma }{3\gamma -1}}\left(
V_{infl}^{\frac{1}{4}}\right) ^{2\frac{3\gamma -2}{3\gamma -1}} \qquad
\left(\gamma \gtrsim 1 \right)
\end{equation}
which is somewhat lower than unity and decreases with $V_{infl}^{\frac{1}{4}}$.

Such an early inflationary stage can be easily incorporated into the hybrid
model \cite{double}. In particular, if we allow field values considerably
larger than unity (e.g. $\left| \varphi _{0}\right| =\left| \sigma _{0}\right|
\gtrsim 10$ for $\beta \simeq 1/35,\ \lambda \simeq 10^{-2}$, $\mu \simeq
1.1\times 10^{-3}$) the original model gives rise
to an early chaotic-type inflationary stage at $\rho= \rho_{0} \sim \rho_{in}$
which takes care of the initial condition problem.

Linde's potential can be easily obtained in the context of global
supersymmetry (SUSY).
Let us consider a model with gauge group $G$ which breaks spontaneously
at a scale $M.$ The symmetry breaking of $G$ is achieved through a
superpotential which includes the terms \cite{dss}
\begin{equation}
W=S(-\mu ^{2}+\lambda \Phi \bar{\Phi}).
\end{equation}
Here $\Phi ,\bar{\Phi}$ is a conjugate pair of left-handed superfields which
belong to non-trivial representations of $G$ and break it by their vacuum
expectation values (vevs), $S$ is a gauge singlet left-handed superfield,
$\mu $ is a mass scale related to $M$ and $\lambda $ a real and positive
coupling constant. The superpotential terms in $W$ are the dominant
couplings involving the superfields $S$, $\Phi $, $\bar{\Phi}$ which are
consistent with a continuous R-symmetry under which $W\to e^{i\vartheta }W$, 
$S\to e^{i\vartheta }S$, $\Phi\bar{\Phi} \to \Phi\bar{\Phi} $.
The potential obtained from $W$ is 
\begin{equation}
V=\left| -\mu ^{2}+\lambda \Phi \bar{\Phi}\right| ^{2}+\lambda^{2}\left|S\right|
^{2}(\left| \Phi \right| ^{2}+\left| \bar{\Phi}\right| ^{2})+D-terms,
\end{equation}
where the scalar components of the superfields are denoted by the same
symbols as the corresponding superfields. The SUSY minimum $S=0,$ 
$\Phi \bar{\Phi}=$ $\mu ^{2}/\lambda, $
$ \left| \Phi \right| =\left| \bar{\Phi}
\right| $ lies on the D-flat direction $\Phi =\bar{\Phi}^{*}$. By
appropriate gauge and R-transformations on this D-flat direction we can bring
the complex $\Phi $, $\bar{\Phi}$, $S$ fields on the real axis, i.e.
$\Phi =\bar{\Phi}\equiv \frac{1}{2}\varphi$, $S\equiv \frac{1}{\sqrt{2}}\sigma$,
where $\varphi$ and $\sigma$ are real scalar fields. The
potential then becomes 
\begin{equation}
V(\varphi ,\sigma )=(-\mu ^{2}+\frac{1}{4}\lambda \varphi ^{2})^{2}+\frac{1}{
4}\lambda ^{2}\varphi ^{2}\sigma ^{2}.
\end{equation}
This is Linde's potential (with $\lambda _{1}=\lambda ^{2}$) apart from a
mass-squared term for $\sigma$. A tiny $m_{\sigma}^{2}$ can be
generated as a result of soft SUSY-breaking or a larger one
due to the promotion of global SUSY to local as we will see shortly.
In the absence of $m_{\sigma}^{2}$ the necessary slope $V^{^{\prime}}$
could be provided by radiative corrections \cite{dss}.

Supersymmetry cannot, of course, remain just global. Thus, we must at some
point face the problem of extending the hybrid model to incorporate
supergravity. We might naively have thought that we could evade the
complications of supergravity by staying at small energies and field values.
Unfortunately, there are two reasons for which this is not possible.
Firstly, as our earlier discussion made it clear, the problem of initial
conditions by definition cannot be addressed at small field values and energies. 
A second very well-known reason
is that, in the case that the potential during inflation is dominated by the
F-term, supergravity tends to give a large mass to almost all fields,
thereby eliminating most candidate inflatons \cite{lyth,st}.
This can be easily seen by considering the F-term potential in supergravity 
\begin{equation}
V_{F}=e^{K}(\cdots ),
\end{equation}
where $K$ is the K\"ahler potential. Let us assume that our candidate
inflaton field $S$ is canonically normalized for $\left| S\right| ^{2}\ll 1$
and the K\"ahler potential admits an expansion $K=\left| S\right|
^{2}+\ldots $ . Then, 
\begin{equation}
m_{S}^{2}=\frac{\partial ^{2}K}{\partial S\partial S^{*}}V_{F}+\ldots
=\left( 1+\ldots \right) V_{F}+\ldots=V_{F}+\ldots \ .
\end{equation}
Thus, during inflation, no matter how small $V_{infl}\simeq V_{F_{infl}}$
is, there is always a contribution to $m_{S}^{2}$  $\simeq V_{infl}$ or
a contribution $\simeq 1$ to the ``slow-roll'' parameter $\left|\eta\right|.$
There could very well exist other contributions to $\eta $ partially
cancelling the one just described but their existence will depend on the
details of the model. Therefore it seems that in the context of supergravity
it is easy to add to the potential of the hybrid model a sizeable
mass-squared term for the inflaton $\sigma$. We only have to understand why
$\beta\equiv \frac{m_{\sigma}^{2}}{\mu ^{4}}$ is not of order unity but much
smaller.

In order to investigate the effect of supergravity on the simple globally
supersymmetric hybrid model discussed above we restrict ourselves to the
inflationary trajectory ($\Phi $ $=\bar{\Phi}=0$) and use the simple
superpotential 
\begin{equation}
W=-\mu ^{2}S
\end{equation}
involving just the gauge singlet superfield $S.$ If the minimal K\"ahler
potential $K=\left| S\right| ^{2}\ $leading to canonical kinetic terms
for $\sigma $ is employed the ``canonical'' potential $V_{can}$ acquires a
slope and becomes \cite{lyth,pan,linde97}
\begin{equation}
V_{can}=\mu ^{4}\left( 1-\left| S\right| ^{2}\ +\left| S\right| ^{4}\right) {
e^{\left| S\right| ^{2}\ }=\mu }^{4}\sum_{k=0}^{\infty }\frac{(k-1)^{2}}{k!}
\left| S\right| ^{2k}.
\end{equation}
Obviously $V_{can}$ does not allow inflation unless $\left| S\right| ^{2}\
\ll 1.$ From its expansion as a power series in $\left| S\right| ^{2}\ $we
see that, due to an ``accidental'' cancellation, the linear term in $\left|
S\right| ^{2}\ $ is missing and no mass-squared term is generated for $\sigma$.
Small deviations from the minimal form of the K\"ahler potential
respecting the R-symmetry lead to a K\"ahler potential \cite{pan1} 
\begin{equation}
K=\left| S\right| ^{2}\ -\frac{\alpha }{4}\left| S\right| ^{4}+\ldots \qquad
\left( \left| S\right| ^{2}\ <<1\right).
\end{equation}
This, in turn, gives rise to a potential admitting an expansion 
\begin{equation}
V=\mu ^{4}\left( 1+\alpha \left| S\right| ^{2}\ +\ldots \right) \qquad
\left( \left| S\right| ^{2}\ <<1\right)
\end{equation}
in which a linear term in $\left| S\right| ^{2}\ $proportional to the small
parameter $\alpha $ is now generated. All higher powers of $\left| S\right|
^{2}\ $are still present in the series with coefficients only slightly
different from the corresponding ones in the expansion of $V_{can}$.

The above discussion seems to indicate that the only potential source of
mass for $\sigma $ is the next to leading term in the expansion of the K\"ahler
potential in powers of $\left| S\right| ^{2}\ $which must have
a small and negative coefficient. This conclusion is certainly correct if
all other fields are assumed to play absolutely no role during inflation.
There could exist fields, however, which do not contribute to the
superpotential and are $G-$singlets, but do contribute to the mass-squared
of $\sigma $ if they acquire large vevs. Such fields could destroy the
``miraculous'' cancellation leading to a massless $\sigma $ in the
case of the minimal K\"ahler potential \cite{lr} but could also
generate new ``miraculous'' cancellations if other types of possibly better
motivated K\"ahler potentials with $\alpha <0$ are employed
\cite{costas}.

Let us consider a $G$-singlet chiral superfield $Z$ which does not
contribute to the superpotential at all because, for instance, it has
non-zero charge, let us say $-1,$ under an ``anomalous'' $U(1)$ gauge
symmetry and, as we assume, all other superfields which have a $U(1)$ charge
can be safely ignored. Also let us assume that $K=K_{1}(\left| S\right|
^{2})+K_{2}(\left| Z\right| ^{2}).$ Then, with the parameters $\mu $ and
$\lambda $ in $W$ renamed as $\mu ^{^{\prime }}$ and $\lambda ^{^{\prime }}$,
the scalar potential (always with $\Phi $ $=\bar{\Phi}=0$) becomes 
\begin{eqnarray}
V=\mu ^{^{\prime }4}\left\{ \left| 1+\frac{\partial K}{\partial S}S\right|
^{2}\left( \frac{\partial ^{2}K}{\partial S\partial S^{*}}\right)
^{-1}+ \left| \frac{\partial K}{\partial Z}S\right| ^{2}\left( \frac{
\partial ^{2}K}{\partial Z\partial Z^{*}}\right) ^{-1}-3\left|S\right| ^{2}
\right\} e^{K}  \nonumber
\end{eqnarray}
\begin{equation}
+\frac{1}{2}g_{1}^{2}\left( \frac{\partial K}{ \partial Z}Z-\xi \right) ^{2},
\end{equation}
where the first(second) term is the F(D)-term, $\xi >0$ is a
Fayet-Iliopoulos term and $g_{1}$ the gauge coupling of the ``anomalous''
$U(1)$ gauge symmetry. Minimization of such a potential for fixed
$\left|S\right| ^{2}$ not much larger than unity, assuming
$\left| S\right| ^{2}\ $ takes values away from any points where the potential
is singular and $\mu^{^{\prime }2}\ll \xi $, typically gives rise to a
$<\left| Z\right|^{2}>\equiv v^{2}\sim \xi $ with
$\left( \left| \frac{\partial K}{\partial Z}\right| ^{2}
\left( \frac{\partial ^{2}K}{\partial Z\partial Z^{*}}\right)^{-1}
\right) _{\left| Z\right| =v}\sim v^{2}\sim \xi $ and a contribution
to the mass-squared of $\sigma$ 
\begin{equation}
\delta m_{\sigma }^{2}=\left( \left| \frac{\partial K}{\partial Z}\right|
^{2}\left( \frac{\partial ^{2}K}{\partial Z\partial Z^{*}}\right)
^{-1}\right) _{\left| Z\right| =v}\mu ^{^{\prime }4}e^{K_{2}(v^{2})}
\end{equation}
of the order of $\xi $ in units of the false vacuum energy density. For the
sake of convenience we absorb the factor $e^{K_{2}(v^{2})}$ appearing in the
F-term potential in the reintroduced parameters $\mu =\mu ^{^{\prime
}}e^{K_{2}(v^{2})/4}$ and $\lambda =\lambda ^{^{\prime }}e^{K_{2}(v^{2})/2}$
obeying the relation $\frac{\mu }{\sqrt{\lambda }}=\frac{\mu ^{^{\prime }}}
{\sqrt{\lambda ^{^{\prime }}}}$.

Notice that the contribution of $Z$ to $m_{\sigma }^{2}$ is positive.
Therefore to make use of the above discussion we should find K\"ahler
potentials $K_{1}(\left| S\right| ^{2})$ whose expansion in powers of
$\left| S\right| ^{2}$ has a positive next to leading term (i.e. $\alpha <0$).
A class of such K\"ahler potentials is given by 
\begin{equation}
K_{1}(\left| S\right| ^{2})=-N\ln \left( 1-\frac{\left| S\right| ^{2}}{N}
\right) \qquad \left( \left| S\right| ^{2}<N\right) ,
\end{equation}
where $N$ is an integer. The corresponding K\"ahler manifold is the
coset space $SU(1,1)/U(1)$ with constant scalar curvature $2/N.$ Expanding
$K_{1}$ in powers of $\left| S\right| ^{2}$ we see that $\alpha =-2/N$
and therefore 
\begin{equation}
m_{\sigma }^{2}=\left( -\frac{2}{N}+\left| \frac{\partial K}{\partial Z}
\right| ^{2}\left( \frac{\partial ^{2}K}{\partial Z\partial Z^{*}}\right)
^{-1}\right) _{\left| Z\right| =v}\mu ^{4}\equiv \beta \mu ^{4}.
\end{equation}
For all $N$ we can make $m_{\sigma }^{2}$ positive (or, by fine tuning,
zero) through appropriately chosen vevs ($\xi $ parameters) of $Z$-type
fields.

It would be very interesting if the contribution of $Z$ to the mass-squared
of $\sigma $ in units of the false vacuum energy density were independent of
the value of $Z$. This is exactly the case if $Z$ enters the K\"ahler
potential through a function $K_{2}$ of the ``no-scale'' type 
\begin{equation}
K_{2}(\left| Z\right| ^{2})=-n\ln \left( -\ln \left| Z\right| ^{2}\right)
\qquad \left( 0<\left| Z\right| ^{2}<1\right),
\end{equation}
where $n$ is an integer. The corresponding K\"ahler manifold is again
the coset space $SU(1,1)/U(1)$ with constant scalar curvature $2/n$. Such a
choice makes the contribution $\delta m_{\sigma }^{2}$ of $Z$ to $m_{\sigma
}^{2}$ an integer multiple of $\mu ^{4},$
namely $\delta m_{\sigma }^{2}=n\mu ^{4}$. With this choice of $K_{2}$ we
obtain 
\begin{equation}
m_{\sigma }^{2}=\left( -\frac{2}{N}+n\right) \mu ^{4}.
\end{equation}
Obviously the most interesting cases occur for $N=1$ or $N=2$ because $2/N$
is an integer and the option of naturally making $m_{\sigma }^{2}$ vanish
for $n=2$ or $n=1,$ respectively becomes now available. A small positive
$m_{\sigma }^{2}$ could be subsequently generated through additional $Z$-type
fields which acquire vevs of the order of appropriately chosen $\xi $
parameters.

The choices $N=1$ or $N=2$ deserve particular attention for the additional
reason that in these cases all supergravity corrections to the F-term potential
are proportional to the mass-squared $m_{\sigma }^{2}$ of the field $\sigma$
or, equivalently, to the parameter $\beta .$ This offers the possibility of
suppressing or even eliminating all supergravity corrections to the
inflationary trajectory by suppressing or making vanish the parameter $\beta$.
Indeed, substituting the K\"ahler potential $K_{1}(\left| S\right| ^{2})$
of Eq. (18) in Eq. (16) and minimizing with respect to $Z$
at fixed $\left| S\right| ^{2}$ we obtain for $N=1,2$ only 
\begin{equation}
V\simeq \mu ^{4}\left\{ 1+\beta \left| S\right| ^{2}\ \left( 1-\frac{\left|
S\right| ^{2}}{N}\right) ^{-N}\right\} \qquad \left( N=1,2\right) 
\end{equation} 
(up to terms $\sim \mu ^{8}$) independently of the mechanism chosen to make
$\beta \geq 0.$ Such models allow for inflation at inflaton field values
close to $1$ or even slightly larger and lead to a possibly detectable
$\left( \frac{\Delta T}{T}\right) _{T}$ \cite{costas}. For relatively small
$\left|S\right| ^{2}$ we obtain the original hybrid model. In particular, with
the choice of $K_{2}(\left| Z\right| ^{2})$ of Eq. (20) the combinations
$(N, n)=(1, 2)$ and $(N, n)=(2, 1)$ give $\beta =0$ and consequently a completely
flat potential. [As already mentioned a small $\beta$ could be generated through
additional $Z$-type fields.] These models with $\beta=0$ could be
regarded as a justification for the SUSY hybrid inflationary scenario
\cite{dss} in which supergravity is neglected completely and the necessary slope
$V^{^{\prime}}$ is provided entirely by radiative corrections.  

Let us now discuss the initial conditions in a model with $\beta =0$ and a
classically completely flat inflationary trajectory.
Our specific model involves, in addition to the superfields $S$, $\Phi $,
$\bar{\Phi}$, one $G$-singlet superfields $Z$ with charge $-1$ under the
``anomalous'' $U(1)$ gauge symmetry. The K\"ahler potential is chosen
to be 
\begin{equation}
K=-\ln \left( 1-\left| S\right| ^{2}\right) -2\ln \left( -\ln \left|
Z\right| ^{2}\right) +\left| \Phi \right| ^{2}+\left| \bar{\Phi}\right| ^{2}
\end{equation}
($\left| S\right| ^{2}<1, 0<\left| Z\right| ^{2}<1$) with the
superpotential always being given by
$W=S(-\mu ^{^{\prime }2}+\lambda ^{^{\prime }}\Phi \bar{\Phi}).$
We define the canonically normalized real scalar fields $\sigma_{infl}$
and $\zeta $ through the relations 
\begin{equation}
\tanh \frac{\sigma _{infl}}{\sqrt{2}}\equiv ReS,\ \ e^{-\zeta }\equiv
-\frac{\xi }{2}\ln \left| Z\right| ^{2},
\end{equation}
with the complex scalar fields $S,\ Z$ brought to the real axis by symmetry
transformations. To simplify the discussion we further set $\Phi =\bar{\Phi}=
\frac{\varphi }{2}$, where $\varphi $ is a canonically normalized real scalar
field, and we consider a trancated version of the complete scalar potential 
\begin{equation}
V=\frac{\lambda ^{2}}{4}\varphi ^{2}\left( \cosh \left( \sqrt{2}\sigma
_{infl}\right) -1\right) e^{2\zeta }+\frac{1}{2}g_{1}^{2}\xi ^{2}\left(
e^{\zeta }-1\right) ^{2}
\end{equation}
(technically justified for $\frac{\mu ^{2}}{\lambda }\ll \varphi
^{2}\lesssim 10^{-1}$) possessing all its salient features ($\lambda
=\lambda ^{^{\prime }}\frac{\xi }{2}$ and $\mu ^{4}=\mu ^{^{\prime }4}\frac{
\xi ^{2}}{4}$ such that $\frac{\mu }{\sqrt{\lambda }}=\frac{\mu ^{^{\prime }}
}{\sqrt{\lambda ^{^{\prime }}}}$). We assume that initially $\left| \sigma
_{infl_{0}}\right| \gg 1$, $e^{\zeta _{0}}\ll 1$, $\varphi _{0}^{2}\sim
10^{-1}$ and the initial time derivatives of all fields vanish. Notice that $
e^{\zeta _{0}}\ll 1$ is required in order for $\rho _{0}\lesssim 1$ if $
\left| \sigma _{infl_{0}}\right| $ is sufficiently large. Then, $e^{\zeta }$
starts decreasing further unless the F-term of $V$ is smaller than $\rho
_{0}e^{\zeta _{0}}$ to begin with. To ensure a sufficiently fast decrease of 
$\varphi ^{2}$ we assume that $\frac{\partial ^{2}V}{\partial \varphi ^{2}}
\gtrsim \rho $ holds from the beginning which, for the initial conditions
adopted, translates into 
\begin{equation}
\left( \cosh \left( \sqrt{2}\sigma _{infl_{0}}\right) -1\right) e^{2\zeta
_{0}}\gtrsim \frac{g_{1}^{2}\xi ^{2}}{\lambda ^{2}}.
\end{equation}
With $\varphi ^{2}$ decreasing fast the relation $\frac{1}{V}\frac{\partial V
}{\partial \zeta }\simeq \frac{1}{V}\frac{\partial ^{2}V}{\partial \zeta ^{2}
}\simeq -2e^{\zeta }$ ($e^{\zeta }\ll 1$) is soon established and the
universe experiences a stage of ``chaotic'' D-term inflation with
$H=H_{1}\simeq \frac{1}{\sqrt{6}}g_{1}\xi (1-e^{\zeta })$ which begins when
$\zeta =\zeta _{beg}\lesssim \zeta _{0}<0.$ The total number of e-foldings
$N_{tot}$ as $\zeta $ varies from $\zeta _{beg}$ towards its minimum at
$\zeta _{\min }\simeq 0$ is 
\begin{equation}
N_{tot}\gtrsim \frac{1}{2}\left( e^{-\zeta _{0}}-e\right) 
\end{equation}
(assuming inflation ends at $\zeta _{end}=-1$). Moreover, $\frac{\partial
^{2}V}{\partial \sigma _{infl}^{2}}\simeq \varphi ^{2}\frac{\partial ^{2}V}{
\partial \varphi ^{2}}.$ Consequently, even if initially $\frac{\partial
^{2}V}{\partial \sigma _{infl}^{2}}\gtrsim \rho $ (i.e. $\left| \sigma
_{infl_{0}}\right| \gg 1$), very soon $\frac{\partial ^{2}V}{\partial \sigma
_{infl}^{2}}\ll \rho $ and $\left| \sigma _{infl}\right| $ stays large with
$\varphi ^{2}$ becoming very small. Thus, when the ``chaotic'' D-term
inflation is over the field configuration is close to the inflationary
trajectory but $\sigma _{infl}$ does not reach its terminal velocity as long
as $\rho $ is dominated by the coherent oscillations of the massive
field $\zeta$ about its minimum.
Actually, even if the initial field values violate the condition in Eq. (26)
and the field configuration fails to approach the inflationary trajectory
during the ``chaotic'' D-term inflation
it may still succeed in approaching it during the period in which $\rho$ is
dominated by the oscillating field $\zeta$.
The ``observable'' inflation starts only after $\rho \sim \mu ^{4}.$ 

A numerical investigation of the complete potential reveals the
existence of more natural initial conditions than the above
simplified analysis indicates. To provide an example in our model
with classically flat inflationary trajectory we consider the choice
$\mu=2.485\times 10^{-4}, \lambda=4\times 10^{-3}$
obtainable \cite{laz} in the SUSY hybrid inflation \cite{dss}.
We also choose $g_{1}=\frac{1}{\sqrt{2}}, \xi=\frac{1}{\sqrt{12}}$ and
$\Phi $ $=\bar{\Phi}=\frac{1}{2}(\varphi $ $+i\psi )$ (along a D-flat direction),
where $\varphi $, $\psi $ are canonically normalized real scalar fields.
Then, it is possible to start at $\rho_{0}\simeq 0.0176$ with
$\varphi _{0}=\psi_{0}\leq\sqrt{2}$ (or $\varphi _{0}\leq 2 $, $\psi_{0}=0$),
$\sigma _{infl_{0}}=1.7$, $\zeta _{0}=-2.5$ 
and zero initial time derivatives for all fields.
An alternative possibility with $\rho_{0}\simeq 1$ 
is to set $\varphi _{0}=\psi_{0}=2.2$ (or $\varphi _{0}=3.1 $, $\psi_{0}=0$),
$\zeta _{0}=-2.1$, $\sigma _{infl_{0}}=5$, 
$(\frac{d}{dt}\sigma _{infl})_{0}=-1$ and assume that the initial time
derivatives for the remaining fields vanish.
Thus, our scenario allows for a quite natural starting point
involving field values which are neither very small nor very large and
an initial energy density $\rho_{0} \sim 1$ possibly equally
partitioned into kinetic and potential.

In summary, hybrid inflation is a natural scenario in the absence of
supersymmetry. In the context of supergravity, however, it has to face two
potential problems. These are the suppression of the inflaton mass
and the implementation of a mechanism providing reasonable initial
conditions. Both problems can be solved in a class of models involving
K\"ahler potentials associated with products of $SU(1,1)/U(1)$ K\"ahler
manifolds and ``decoupled'' fields acquiring large vevs through D-terms.

\section*{}
This research was supported in part by EU under TMR contract
ERBFMRX-CT96-0090.

\def\npb#1#2#3{(#1),~{\it Nucl. Phys.}~{\bf B~#2},~#3}
\def\pl#1#2#3{(#1),~{\it Phys. Lett.}~{\bf #2~B},~#3}
\def\plb#1#2#3{(#1),~{\it Phys. Lett.}~{\bf B~#2},~#3}
\def\pr#1#2#3{(#1),~{\it Phys. Reports}~{\bf #2},~#3}
\def\prd#1#2#3{(#1),~{\it Phys. Rev.}~{\bf D~#2},~#3}
\def\prl#1#2#3{(#1),~{\it Phys. Rev. Lett.}~{\bf #2},~#3}
\def\ibid#1#2#3{(#1),~{\it ibid.}~{\bf ~#2},~#3}
\def\apjl#1#2#3{(#1),~{\it Astrophys. J. Lett.}~{\bf #2},~#3}

\end{document}